\documentclass[12 pt, draft, onecolumn]{IEEEtran}  

\IEEEoverridecommandlockouts                              

\newcommand{\upperRomannumeral}[1]{\uppercase\expandafter{\romannumeral#1}}

\usepackage{color}
\usepackage[margin=0.75in]{geometry}
\usepackage{algorithm}
\usepackage[noend]{algpseudocode}
\usepackage{cite}
\usepackage{subcaption}
\usepackage{enumitem}
\usepackage{caption}
\usepackage{graphicx} 
\usepackage{amsmath, amsfonts, bm} 
\usepackage{amssymb}  
\usepackage[normalem]{ulem}
\usepackage{bbm}
\setkeys{Gin}{draft=false}
\usepackage{ntheorem}
\newtheorem*{remark}{Remark}
\begin{document}
\bstctlcite{IEEEexample:BSTcontrol}
\title{
An Application of Storage-Optimal MatDot Codes for Coded Matrix Multiplication: Fast k-Nearest Neighbors Estimation
}


\author{Utsav Sheth$^{1}$, Sanghamitra Dutta$^{1}$, Malhar Chaudhari$^{1}$, Haewon Jeong$^{1}$, Yaoqing Yang$^{1}$, Jukka Kohonen$^{2}$, Teemu Roos$^{2}$, Pulkit Grover$^{1}$  
\thanks{$^{1}$Carnegie Mellon University, Pittsburgh, USA. Corresponding Author Contact:  {\tt\small  pulkit@cmu.edu}.}%
\thanks{$^{2}$University of Helsinki, Finland. Corresponding Author Contact:
        {\tt\small teemu.roos@cs.helsinki.fi}.}%
\thanks{This work is accepted as a short paper at the IEEE Big Data 2018~\cite{seth2018bigdata}.}
}%
\maketitle
\thispagestyle{empty}
\pagestyle{empty}

\begin{abstract}
We propose a novel application of coded computing to the problem of the nearest neighbor estimation using MatDot Codes~\cite{fahim2017optimal} that are known to be optimal for matrix multiplication in terms of recovery threshold under storage constraints. In approximate nearest neighbor algorithms, it is common to construct efficient in-memory indexes to improve query response time. One such strategy is Multiple Random Projection Trees (MRPT), which reduces the set of candidate points over which Euclidean distance calculations are performed. However, this may result in a high memory footprint and possibly paging penalties for large or high-dimensional data. Here we propose two techniques to parallelize MRPT that exploit data and model parallelism respectively by dividing both the data storage and the computation efforts among different nodes in a distributed computing cluster. This is especially critical when a single compute node cannot hold the complete dataset in memory. We also propose a novel coded computation strategy based on MatDot codes for the model-parallel architecture that, in a straggler-prone environment, achieves the storage-optimal recovery threshold, \textit{i.e.}, the number of nodes that are required to  
serve a query. We experimentally demonstrate that, in the absence of straggling, our distributed approaches require less query time than execution on a single processing node, providing near-linear speedups with respect to the number of worker nodes. Through our experiments on real systems with simulated straggling, we also show that our strategy achieves a faster query execution than the uncoded strategy in a straggler-prone environment. 
\end{abstract}

\section{INTRODUCTION}
We consider the problem of finding the \textit{k} nearest neighbors of a query point in a given high-dimensional dataset. To solve this problem efficiently, our goal is to speed up an existing algorithm~\cite{Hyvonen2016} by parallelizing it, and to make it resilient to \emph{stragglers}~\cite{dean2013tail}. The \textit{k}-nearest neighbor (\textit{k}-NN) problem is often a first step used in a variety of real world applications including genomics~\cite{cancer_knn}, personalized search~\cite{mobile_search_knn}, network security~\cite{intrusion_knn}, and web based recommendation systems~\cite{recommender_knn}.

In the era of Big Data, \textit{k}-NN algorithms are often a bottleneck, as data and dimensionalities grow~\cite{xiong2010data}. There is a rich body of work on fast nearest neighbor retrieval. The existing work can be broadly classified into three categories. The first category of methods speeds up \textit{k}-NN retrieval by reducing the space over which exact distance calculations are performed, by employing space partitioning data structures~\cite{indyk1998approximate,cen2015improved,1265868}. The second category of techniques improves retrieval times by system level parallelism~\cite{liang2009cuknn,huang2012implementation}. Algorithms in the first and second categories are seldom scalable as they require that the whole dataset is held in (shared) memory for optimal performance. For large high-dimensional datasets, marshaling enough resources on a single system is challenging. The third category of methods aims to overcome this challenge by parallelizing storage and computation in a distributed setting. PANDA~\cite{patwary2016panda} and DSI sharding~\cite{yu2015scalable} are examples of data parallel implementations of distributed \textit{k}-NN algorithms. While these techniques rely on specialized or high-performance hardware (e.g. Edison supercomputer for PANDA), the general trend in the distributed systems has been to use general purpose commodity systems~\cite{shvachko2010hadoop}~\cite{barroso2003web}. These approaches also do not address a more serious issue --- the effects of node failures and slow nodes, or ``stragglers''. Schroeder and Gibson~\cite{schroeder2010large} observed as many as $1{,}159$ system failures per year at the Los Alamos National Laboratory. Dean et al.~\cite{dean2013tail} study a real Google service and observe that the slowest 5\% of requests are responsible for half of the total 99$^{th}$ percentile latency. 
%
%

Recently ``Coded Computing''~\cite{NewsletterPaper,gauri2014delay,gauri2015straggler,gauri2014efficient,lee2016speeding,lee2018speeding,dutta2016short,dutta2017coded,fahim2017optimal,dutta2018optimal,dutta2018DNN1,dutta2018DNN2,lee2017matrix,yu2017polynomial,GC1,GC2,yu2018entangled,li2015coded,GC3,azian2017consensus,yang2016logistic,yang2017encoded,yang2016convolution,yang2018ISIT,yang2016encoded,yang2017NIPS,Salman1,Salman2,Salman3,Salman4,Emina1,Emina2,Virtualization,heterogeneousclusters,GC4,Suhas1,Suhas2,Ramtin1,lee2017multicore,jeongFFT,baharav2018straggler,suh2017matrix,mallick2018rateless, wang2018coded, wang2018fundamental,severinson2018block,ye2018communication,haddadpour2018codes,haddadpour2018straggler,ferdinand2016anytime,ferdinand2018hierarchical,song2017pliable,kosaian2018learning,seth2018bigdata,jeong2018locally} has been found to be very useful in combating stragglers and faults, by the efficient use of novel erasure-codes to create redundancy in computing. In this paper, we use one such coded computing technique called \emph{MatDot codes}~\cite{fahim2017optimal,dutta2018optimal} to speed up an approximate $k$-NN algorithm called \textit{Multiple Random Projection Trees} (MRPT)~\cite{Hyvonen2016}. 

The problem of distributed matrix multiplication has always been of significant interest in the coded computing community~\cite{yang2017encoded,lee2018speeding,dutta2016short,lee2017matrix,yu2017polynomial,fahim2017optimal,dutta2018optimal,dutta2018DNN2,yu2018entangled}. In our prior work~\cite{fahim2017optimal,dutta2018optimal}, we proposed a novel coding technique called MatDot codes for the multiplication of two matrices (e.g. $\mathbf{X}^T \mathbf{Q}$) that outperform the recent Polynomial Codes~\cite{yu2017polynomial} with respect to recovery threshold, \emph{i.e.}, the number of worker nodes needed to wait for, out of the total nodes, under storage constraints. More specifically, for the problem of distributed matrix multiplication under the storage constraint that each node can only store a fixed $\frac{1}{m}$ fraction of each matrix, MatDot codes achieve the optimal recovery threshold of $2m-1$~\cite{fahim2017optimal,dutta2018optimal} as compared to Polynomial codes that have a recovery threshold of $m^2$. 

We note that for matrix multiplication $\mathbf{X}^T\mathbf{Q}$ or matrix-vector product $\mathbf{X}^T\mathbf{q}$ under the constraint that only $\frac{1}{m}$ fraction of each operand can be stored at each node, MatDot codes use vertical block-partitioning of the first matrix $\mathbf{X}^T$ as compared to other existing strategies that use either horizontal partition or a combination of both (see \cite{dutta2018DNN2} for discussion on partitioning). Interestingly, as it turns out, the vertical partitioning of the first matrix is better suited for the coded MRPT problem formulation. This is because we are required to perform the multiplication $\mathbf{X}(S)^T\mathbf{Q}$ or $\mathbf{X}(S)^T\mathbf{q}$ in a distributed fashion, where $\mathbf{X}(S)^T$ denotes a sub-matrix of $\mathbf{X}^T$ consisting of the rows of $\mathbf{X}^T$ indexed in a set $S$, such that, $\bm{X}^T$ is known in advance but the set $S$ becomes available only in the online phase (real-time). 

This coded computing problem ensues from our broader goal in this work, which is to speed up MRPT by employing data and model parallelism and straggler-tolerant computing techniques. We differentiate here between \textit{data parallelism} and \textit{model parallelism}. In data parallelism, different nodes process different pieces of data, but each node performs all the computations relevant to the entire model. In model parallelism, the model itself is parallelized across nodes.

MRPT partitions the search space to retrieve approximate \textit{k} nearest neighbors of the query $\mathbf{q}$. It uses a combination of random projection trees and voting to achieve fast queries and high accuracy. In this paper, we propose two enhancements to the MRPT algorithm by parallelizing it in a distributed setting. Our contributions are as follows:
\begin{itemize}[leftmargin=8pt]
\item We propose a distributed implementation of MRPT exploiting data parallelism that experimentally demonstrates faster queries than a single node implementation, even when using CPUs with lower clock speeds. Additionally, the cloud-based virtual machines we use in our experiments for parallel MRPT have a non-zero steal time, \textit{i.e.}, they may be required to wait while others are being served. 


\item We formulate a coded computing problem for MRPT, and then apply coded matrix multiplication strategies, namely MatDot and Systematic MatDot codes to further reduce the query time for the model parallel architecture in a system that is prone to straggling. 
\end{itemize}

The rest of the paper is organized as follows. In Section~\ref{sec:prelim} we explain the MRPT algorithm and describe how it reduces the search-space through projections and voting. In Section~\ref{sec:data_parallel_mrpt_sec} we introduce the Data Parallel Model Implementation of MRPT. In Section~\ref{mrpt_model_parallel} we introduce our proposed model parallel implementation of MRPT and then describe the application of MatDot codes and systematic MatDot codes in our model parallel architecture to achieve a lower recovery threshold~\cite{fahim2017optimal} under straggling. The model parallel architecture is ideal for applications where system components are unreliable and accuracy is important. In Section~\ref{expt} we experimentally demonstrate the advantages of our approach. A conclusion is provided in Section~\ref{sec:conclusion}.

\section{Preliminaries}
\label{sec:prelim}

\subsection{MRPT Algorithm} \label{subsec:mrpt}
This section briefly describes the two stages of the MRPT algorithm: (i) off-line index construction stage and (ii) on-line query stage. Assume that we are given a $d$-dimensional dataset $\mathcal{X}$ consisting of $N$ points, represented as a $d \times N$ matrix $\mathbf{X}$. Given a query point $\mathbf{q}$, the problem of \textit{k}-nearest neighbors involves finding a set of points $\kappa \subseteq \mathcal{X}$ such that $\vert \kappa \vert = k$ and $
dist(\mathbf{x},\mathbf{q}) \leq dist(\mathbf{y},\mathbf{q})
$
for each $\mathbf{x} \in \kappa$, $\mathbf{y} \in \mathcal{X} \backslash \kappa$, and the function $dist(\cdot)$ is the distance function in the \textit{d}-dimensional Euclidean space given by:
\begin{equation}\label{euclidean_distance}
dist(\mathbf{u},\mathbf{v}) = \Vert \mathbf{u}-\mathbf{v}\Vert = \sqrt{\Vert \mathbf{u}\Vert^2 + \Vert \mathbf{v}\Vert^2 - 2\mathbf{u}\cdot \mathbf{v}},
\end{equation}
where $\mathbf{u}$ and $\mathbf{v}$ are two vectors in this space. 

In the MRPT algorithm, a sparse $d$-dimensional random projection vector
$\mathbf{r}$ is chosen, in which each entry $r_i$ is sampled from the following distribution: 
\begin{equation*}
r_i = \begin{cases}
\mathcal{N}(0,1) & \text{with probability}\ a \\
0 & \text{with probability}\ 1-a.
\end{cases}
\end{equation*}
Typically, the sparsity parameter $a$ can be chosen as $\frac{1}{\sqrt{d}}$, as in \cite{Hyvonen2016}, to obtain good accuracy. Now, each $d$-dimensional data-point $p \in \mathcal{X}$ is then projected onto the sparse vector $\mathbf{r}$. The dataset $\mathcal{X}$ is then divided into two subsets at the median point of the projected values. 
The process is then repeated recursively for every subset at a level, with a new random vector $\mathbf{r}$ chosen for that tree level, until depth $\ell$ is reached. Thus, for every tree $t \in \mathcal{T}$, the entire dataset $\mathcal{X}$ is partitioned into $2^\ell$ cells (or leaves), denoted as $L_1,L_2,\ldots,L_{2^l}$, all of which contain $\lceil\frac{N}{2^\ell}\rceil$ or $\lfloor\frac{N}{2^\ell}\rfloor$ data-points.

\subsubsection*{Online Query Stage in MRPT}\label{mrpt_online}
Given a $d$-dimensional query vector $\mathbf{q}$, the first step in the MRPT query stage is to generate a candidate set of indices (pruned data-point indices) $S \subset \{ 1,2,\ldots,N\}$ such that $\vert S\vert$ $\ll N$.

For each Random Projection (RP) tree $t \in$ $\mathcal{T}$, at each level the query vector $\mathbf{q}$ is projected onto the random vector $\mathbf{r}$ for that level and then assigned a branch based on whether its value is greater than or less than the median of the projections of all other data-points with $\mathbf{r}$. This process is then repeated recursively until a leaf is reached. 

Each tree had already partitioned the dataset $\mathcal{X}$ into $2^l$ cells or leaves. For $1 \leq t \leq T$, let $f_t(\cdot)$ be defined as:
\begin{equation}
f_t(\mathbf{x}:\mathbf{q}) = \sum_{i=1}^{2^\ell}\mathbbm{1} (\mathbf{x} \in L_i, \mathbf{q} \in L_i),
\end{equation}
where $\mathbbm{1} (\mathbf{x} \in L_i, \mathbf{q} \in L_i)$ denotes the indicator function that returns $1$ if both $\mathbf{x}$ and $\mathbf{q}$ reside in the same cell. Let $F(\cdot)$ be a function that returns the number of trees in which $\mathbf{x}$ and $\mathbf{q}$ occur in the same leaf, defined as follows:
\begin{equation}
F(\mathbf{x};\mathbf{q}) = \sum_{t=1}^{T}f_t(\mathbf{x},\mathbf{q}).
\end{equation}
The candidate set of indices (pruned points) $S$ can then be finally chosen as follows:
\begin{equation}
S = \{j \subset \{1,2,\ldots,N\} : \mathbf{x}_j \in \mathcal{X} \text{ and }  \textit{F}(\mathbf{x}_j;\mathbf{q}) \geq \nu\}.
\end{equation}
Here, $\nu$ is a pre-configured parameter known as the \textit{voting threshold}. Thus, the set $S$ denotes the set of indices $\subset \{1,2,\ldots,N\}$ for which at least $\nu$ trees have found the corresponding data-point $\mathbf{x}_j$ in the same cell as $\mathbf{q}$.

Finally, exact distance calculations are performed for each  $\mathbf{x}_j$ with $j \in S$, to obtain the approximate \textit{k} nearest neighbors to the query-vector $\mathbf{q}$. The algorithm for this stage is mentioned in Algorithm~\ref{mrpt_algorithm}. Here, \texttt{TREE\_QUERY}$(\mathbf{q},t)$ is a function corresponding to tree $t$ that returns the pruned collection of the indices of the data-points that lie in the same cell (or leaf) as $\mathbf{q}$. Thus,  \texttt{TREE\_QUERY}$(\mathbf{q},t)$ gives $\{j \subset \{1,\ldots,N\}:  \mathbbm{1} (\mathbf{x}_j \in L_i, \mathbf{q} \in L_i) = 1 $ for some $ L_i\}$. The exact distance calculation is discussed again in Section~\ref{mrpt_model_parallel}.

\begin{algorithm}
\caption{The MRPT Query Phase}
\label{mrpt_algorithm}
\begin{algorithmic}[1]
\Procedure{Approximate\_Knn}{q, k, $\mathcal{T}$, $\nu$}
\State \textit{S} $\gets$ $\emptyset$ 
\State \text{Let} \textit{votes} = [0, \ldots, 0] 
\text{be a new $n$-dimensional array}
\For{\textit{t} in  $\mathcal{T}$}
\For{\textit{point} in \texttt{TREE\_QUERY($\mathbf{q}$, \textit{t})}}
\State \textit{votes[point]} $\gets$ \textit{votes[point]} + 1
\If {$\textit{votes}[\textit{point}] = \nu$}
\State \textit{S} $\gets$ \textit{S} $\cup$ \{\textit{point}\}
\EndIf
\EndFor
\EndFor
\State \textbf{return} \texttt{EXACT\_KNN}($\mathbf{q}$, \textit{k}, \textit{S})
\EndProcedure
\end{algorithmic}
\end{algorithm}

\subsection{Coded Matrix Multiplication} \label{subsec:code_comp}
Coded computing combines distributed numerical algorithms and error correcting codes (ECCs) to mitigate unreliable processors and randomness in their response time. In this work, we focus on coded matrix multiplication as matrix multiplication is the main bottleneck in MRPT algorithm. Coded computing has been used extensively for different computation objectives such as neural-network training, FFT, iterative computing, distributed regression, and convolutions (see \cite{NewsletterPaper} for review).

\textbf{System Model: } We want to compute $\bm{C} = \bm{A} \bm{B}$ where $\bm{A}$ and $\bm{B}$ are $N$-by-$N$ matrices. \emph{A master node} distributes the computation to $P$ \emph{worker nodes}. A worker node has limited memory/computing power, so each node can receive the $1/m$-th fraction of matrices $\bm{A}$ and $\bm{B}$. After completing its computation, a worker reports the result to \emph{a fusion node.} 
\emph{Recovery threshold} is defined as the worst-case number of workers needed to recover the final result. 

Let $K$ be the number of workers needed to complete the computation if all worker nodes are reliable ($K$ is different depending on how we split the matrix). In reality, some processors are significantly slower than the others due to queuing delays or random faults in the processor\cite{lee2018speeding}. Without any reliability measure to alleviate straggler problems, computation completion time would be dominated by few stragglers. Our aim is to use more than $K$ worker nodes by adding some redundancies, which are carefully designed by applying the ideas from coding theory, so that the whole computation can be resilient to stragglers.

\textbf{MatDot codes:} The matrix $\mathbf{A}$  is split vertically into $m$ column blocks, and $\mathbf{B}$ is split horizontally into $m$ row blocks:
\begin{equation}\label{eq:matdot_split}
\mathbf{A} = \left[\mathbf{A}_1 \ \mathbf{A}_2 \ \ldots \ \mathbf{A}_{m}\right],\;\;\; \mathbf{B}=\left[\begin{array}c \mathbf{B}_1\\ \vdots \\\mathbf{B}_{m}\end{array}\right],
\end{equation}
where $\mathbf{A}_{i},\mathbf{B}_{i}$ ($i=1,\ldots,m$) are $N \times N/m$ and $N/m \times N$ dimensional submatrices, respectively. 

The matrices $\mathbf{A}$ and $\mathbf{B}$ are then encoded as polynomials:
\begin{equation}
p_{\mathbf{A}}(x) = \sum_{i=1}^{m} \mathbf{A}_i x^{i-1}, \;\;\; p_{\mathbf{B}}(x) = \sum_{j=1}^{m} \mathbf{B}_j x^{m-j}.
\end{equation}
A master node distributes encoded matrices, $p_{\mathbf{A}}(\alpha_i)$ and $p_{\mathbf{B}}(\alpha_i)$ to the $i$-th worker node ($i = 1, \ldots, P$).
Then the $i$-th worker node computes the following product at $x=\alpha_i$:
\begin{equation}
p_{\mathbf{C}}(x) = \sum_{i=1}^{m} \sum_{j=1}^{m} \mathbf{A}_i \mathbf{B}_j x^{m-1+(i-j)},
\end{equation}
and returns the result to the master node. 
Note that the coefficient of $x^{m-1}$ in $p_{\mathbf{C}}(x)$ is $\mathbf{C} = \sum_{i=1}^{m} A_i B_i$. 
Since $p_{\mathbf{C}}(x)$ is a polynomial of degree $2m-2$, its coefficients can be recovered by the master node as soon as it receives the values of $p_{\mathbf{C}}(x)$ at any  $2m-1$  distinct points. Hence the recovery threshold is $K = 2m-1$. 
This is provably the optimal recovery threshold, when a worker node can store $\frac{1}{m}$-th fraction of each input matrix~\cite{fahim2017optimal,dutta2018optimal}. 

\textbf{Systematic MatDot codes:} A code is called \emph{systematic} if, for the first $m$ worker nodes, the output of the $r$-th worker node is the product $\mathbf{A}_{r}\mathbf{B}_{r}$. We refer to the first $m$ worker nodes as \emph{systematic worker nodes}. Having systematic nodes is useful because if all the systematic nodes complete their computation in time, there is no need for decoding. 
Systematic MatDot codes are achieved by applying different encoding polynomials. 
Let $p_\mathbf{A}(x)=\sum_{i=1}^{m} \mathbf{A}_{i} L_{i}(x)$ and $p_\mathbf{B}(x)=\sum_{i=1}^{m} \mathbf{B}_{i} L_{i}(x)$ where $L_i(x)$ is defined as follows for $i \in \{ 1,\ldots, m \}$:
\begin{equation}\label{eq:lpol}
    L_i(x)=\displaystyle\prod\limits_{\substack{  j\in \{1, \ldots, m\} \setminus \{i\}}}\frac{ x-x_j}{x_i-x_j}.
\end{equation}
Using these polynomials, the worst-case recovery threshold remains the same as non-systematic MatDot codes~\cite{fahim2017optimal,dutta2018optimal}.

\begin{remark}[1]
MatDot and Systematic MatDot codes use vertical partitioning of the first matrix which is well suited for the problem of coded MRPT as compared to strategies that use horizontal partitioning or a combination of horizontal and vertical partitioning for the first matrix. This is because in the coded MRPT formulation, we are required to perform the multiplication $\mathbf{X}(S)^T\mathbf{Q}$ or $\mathbf{X}(S)^T\mathbf{q}$ in a distributed fashion, where $\mathbf{X}(S)^T$ denotes a sub-matrix of $\mathbf{X}^T$ consisting of the rows of $\mathbf{X}^T$ indexed in a set $S$, such that the set $S$ is available only in the online phase (real-time). 
\end{remark}

\section{DATA PARALLEL MRPT}\label{sec:data_parallel_mrpt_sec}
\begin{figure}[thpb]
\centering
\includegraphics[width=8.5cm]{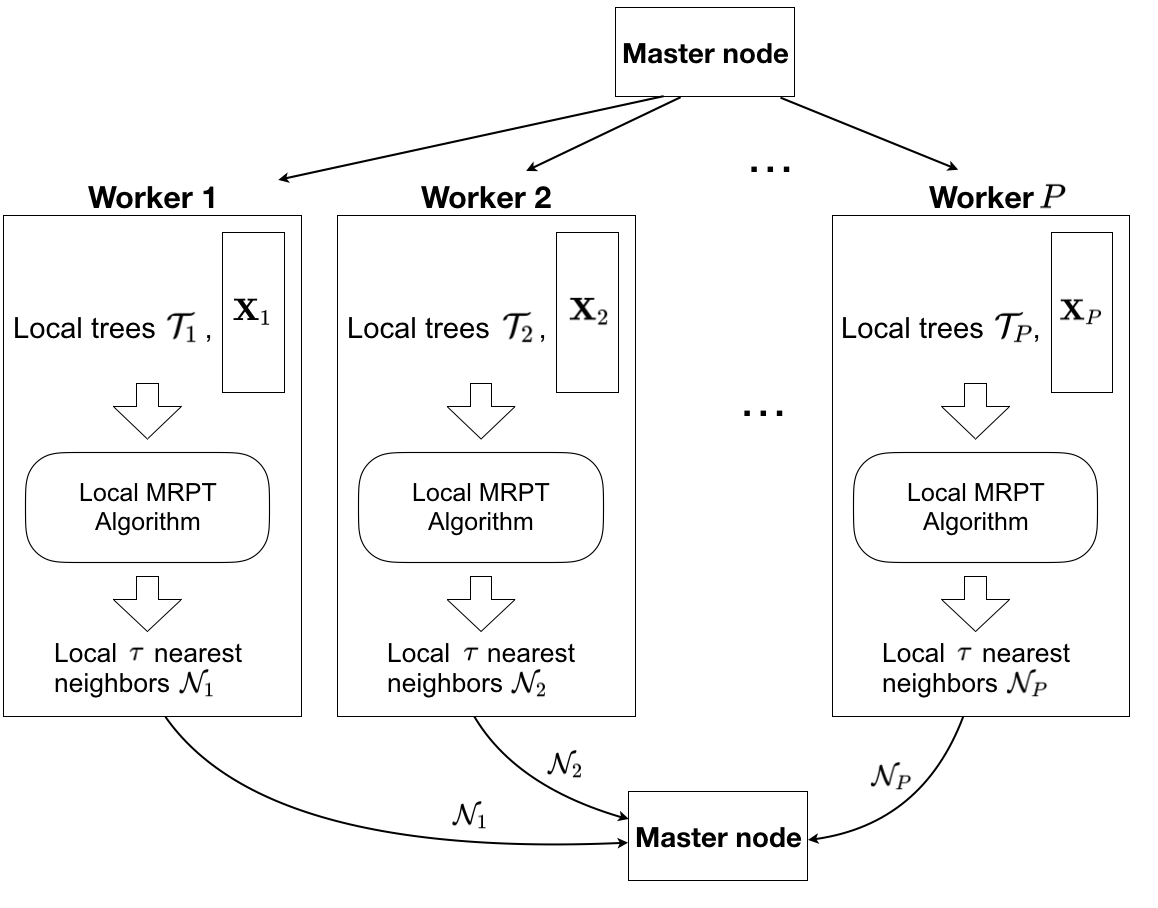}
\caption{Data Parallel MRPT Architecture.}
\label{data_parallel}
\end{figure}

Now, we model MRPT as a problem in data parallelism and describe our first strategy to parallelize the algorithm. Consider a distributed computing cluster having a single \textit{master node} and $P$ \textit{worker nodes} as shown in Fig. \ref{data_parallel}. 

Given a $ d\times N$ matrix $\textbf{X}$ representing the set of data-points $\mathcal{X}$ and a cluster with $P$ worker nodes, we randomly split $\mathbf{X}$ vertically into $P$ disjoint, vertical partitions $\mathbf{X}_i$ for $i \in {\{1,2,\ldots,P\}}$.
 Thus,
$\mathbf{X}= \begin{bmatrix} \mathbf{X}_1 \vert \mathbf{X}_2 \vert \dots \vert \mathbf{X}_P \end{bmatrix},$ where each $\mathbf{X}_i$ is of dimension $d \times \frac{N}{P}.$

We distribute each partition $\mathbf{X}_i$ across the $P$ worker nodes in the cluster such that the worker $W_i$ contains partition $\mathbf{X}_i$. Each worker then runs the MRPT index construction algorithm described in Section~\ref{subsec:mrpt} to build a local MRPT set of trees $\mathcal{T}_i$ from its partition $\mathbf{X}_i$ of $\mathbf{X}$. 

Given a query $\mathbf{q}$, the master node transmits $\mathbf{q}$ to each worker node $W_i$. Then $W_i$ uses the MRPT query algorithm described in Section~\ref{mrpt_online} on its trees $\mathcal{T}_i$ to determine the $\tau$ nearest neighbors of $\mathbf{q}$. We use the same voting threshold $\nu$ in all workers. After local voting, a local exact distance calculation step is performed to narrow down to $\tau$ data-points. Finally, each worker node $W_i$ then returns the indices of its set of $\tau \; (\ge k)$ nearest neighbors \textit{$N_i$}, from the partition $\mathbf{X}_i$ of $\mathbf{X}$, to the master node along with their exact distances from the query $\mathbf{q}$. Then, the master node determines the final set of \textit{k} nearest neighbors to $\mathbf{q}$ from all the $P\tau$ candidate data-point indices received from all the worker nodes, \textit{i.e.}, the indices of all the data-points in the set $\cup_{i=1}^P N_i$, by sorting the data-points based on their exact distances. 

\begin{remark}[2]:
Recall that $\kappa \subseteq \mathcal{X}$ (with $\vert \kappa\vert = k$) is the set of the true \textit{k} nearest neighbors of $\mathbf{q}$. We let $\kappa_i \subseteq \mathbf{\kappa}$ be the set of the true nearest neighbors to $\mathbf{q}$ that lie in worker $W_i$. To achieve maximum accuracy, we must have $\kappa_i \subseteq N_i$ and therefore the value of $\nu$ chosen for the system must be such that $\vert$\textit{$\kappa_i$}$\vert$ is much less than $\tau$. In fact, a higher value of $\tau$ implies higher chance of containing the desired set $\kappa_i$, though it comes with increased communication cost from worker to master and increased computation at the master node.
\end{remark}

\section{MODEL PARALLEL MRPT}\label{mrpt_model_parallel}

In this section we discuss a model parallel architecture for approximate \textit{k} nearest neighbor search using MRPT. We then propose two enhancements to the model parallel architecture that apply coded distributed matrix multiplication techniques that achieve the optimal recovery threshold in a system that is prone to straggling.

\subsection{Problem Formulation for Model Parallel MRPT}\label{new_sec}
Consider the \textit{d}-dimensional dataset $\mathcal{X}$ as before. In the model parallel architecture, given a query $\mathbf{q}$, we first find the possible candidate set of indices (pruned indices) \textit{S} using the recursive algorithm described in Section~\ref{mrpt_online}. Now the search space for the true nearest neighbors $\kappa$ reduces to the set of data-points whose indices are in $S$, \textit{i.e.}, $\kappa \subseteq \{\mathbf{x}_j: j \in S  \} $.

To find the set $\kappa$, we compute the exact Euclidean distance from each data-point $\mathbf{x}_j$ (for $j \in S$) to the query point $\mathbf{q}$. Examining the terms constituting the Euclidean distance in (\ref{euclidean_distance}), the Euclidean norm $\Vert$$\mathbf{x}_j$$\Vert$ for each $\mathbf{x}_j \in \mathcal{X}$ can be precomputed and the same can be done to get $\Vert$$\mathbf{q}$$\Vert$. We must now only compute the dot product $\mathbf{x}_j$ $\cdot$ $\mathbf{q}$ to obtain the Euclidean distances from each $\mathbf{x}_j$ to $\mathbf{q}$.

To do this, we first represent the data-points indexed in the set $\textit{S}$ as a $  d \times |S| $ matrix $\mathbf{X}(S)$ that contains only the data-points $\mathbf{x}_j$ (columns of $\mathbf{X}$) such that $j \in S$. The transpose of this matrix is the $|S| \times d$ matrix  $\mathbf{X}(S)^T$ that essentially denotes all the rows of the matrix $\mathbf{X}^T$ indexed in $S$.



Consider the column vector $\mathbf{w}$ such that $\mathbf{w} = \mathbf{X}(S)^T\mathbf{q}$. 
Note that each element of the vector $\mathbf{w}$ corresponds to the dot-product $\mathbf{x}_j \cdot \mathbf{q}  = \mathbf{x}_j^T \mathbf{q}$ for some $j \in S$.  The Euclidean distance from $\mathbf{x}_j$ to $\mathbf{q}$ can now be determined as all the terms in (\ref{euclidean_distance}) are known to us, which includes the individual norms as well as the dot product $\mathbf{x}_j \cdot \mathbf{q}$. 

The problem thus reduces to the following: \emph{compute the vector $\mathbf{w}=\mathbf{X}(S)^T\mathbf{q}$ in a distributed computing cluster where $\mathbf{X}^T$ is known in advance but the set of indices $S$ become available only in the online phase (real-time)}. Since the computation $\mathbf{w} = \mathbf{X}(S)^T\mathbf{q}$ is the only stage of the algorithm that must be done at runtime (in the online stage) and scales linearly with $d$, we now discuss several scalable strategies that compute vector $\mathbf{w}$ in a distributed setting.
\subsection{Uncoded Distributed Matrix-Vector Multiplication}\label{uncoded}

\begin{figure}
\centering
\smallskip
\smallskip
\includegraphics[height=4.5cm]{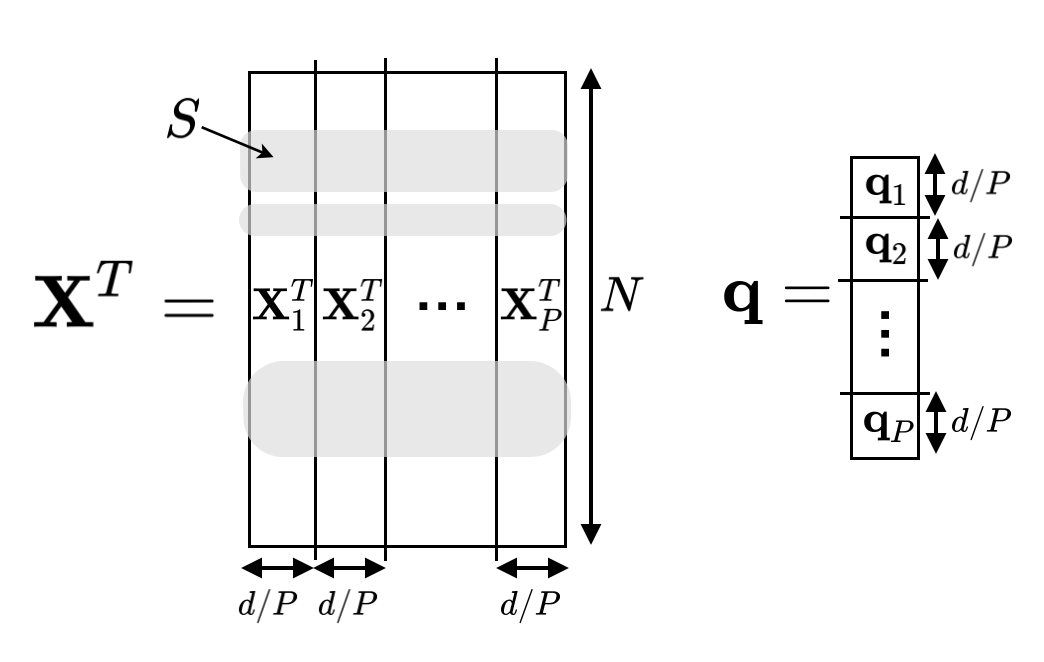}
\caption{Partitioning the data matrix $\mathbf{X}$ and the query vector $\mathbf{q}$ in Model Parallel MRPT.\vspace{-0.2cm}}
\label{model_parallel}
\end{figure}

\begin{figure}
\centering
\smallskip
\includegraphics[height=6.8cm]{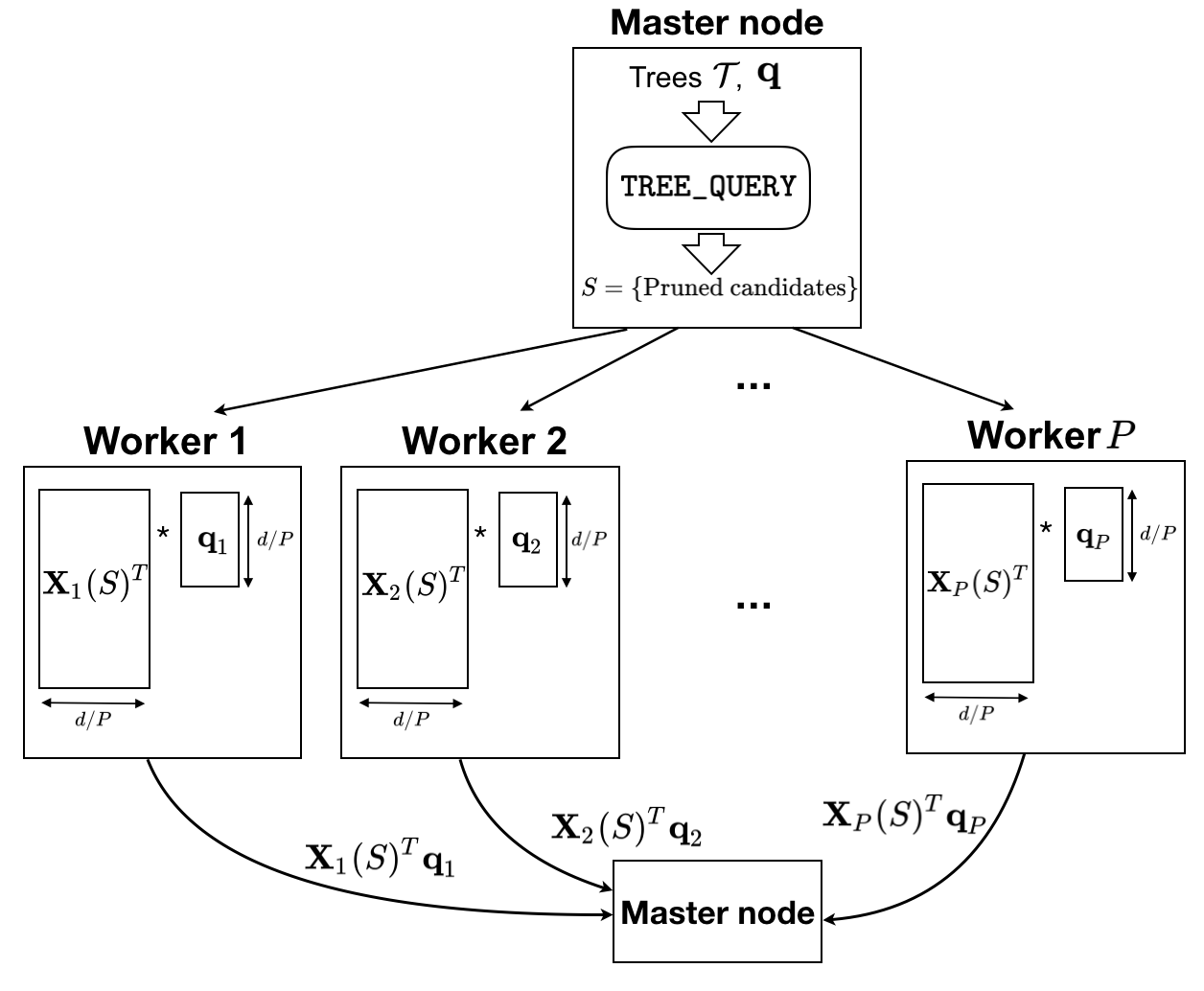}
\caption{Model Parallel Architecture with Uncoded Distributed Matrix Multiplication.}
\label{model_parallel_arch}
\end{figure}

We split $\mathbf{X}^T$ into $P$ equal partitions 
as follows (see Fig.~\ref{model_parallel}):
\begin{equation}
\mathbf{X}^T= \begin{bmatrix}
\mathbf{X}_1^T & \mathbf{X}_2^T & \ldots & \mathbf{X}_P^T
\end{bmatrix}.
\end{equation}
Now consider a cluster consisting of one \textit{master node} and $P$ \textit{worker nodes} as shown in Fig. \ref{model_parallel_arch}. Each partition $\mathbf{X}^T_i$ is distributed across the worker nodes such that worker $\textit{W}_i$ stores the partition ${\mathbf{X}_i}^T$ in advance (off-line). 

Note that, if $\mathbf{X}^T$ is partitioned using the strategy just discussed, the matrix $\mathbf{X}(S)^T$ also gets partitioned as follows:
\begin{equation}
\mathbf{X}(S)^T = \begin{bmatrix}
\mathbf{X}_1(S)^T & \mathbf{X}_2(S)^T & \ldots & \mathbf{X}_P(S)^T.
\end{bmatrix}
\end{equation} 
In the online phase, we only split the query $\mathbf{q}$ into $P$ equal partitions,  $\{\mathbf{q}_i \,:\, i \in \{1,\ldots,P\}\}$ (again see Fig.~\ref{model_parallel}). 
The product $\mathbf{w}$ can then be expressed as:
\begin{equation}\label{uncoded_equation}
\mathbf{w} = \sum_{i=1}^{P}\mathbf{X}_i(S)^T\mathbf{q}_i.
\end{equation}
Given a query $\mathbf{q}$ for which the \textit{k} nearest neighbors must be determined, the master node first computes the possible candidate set \textit{S} for $\mathbf{q}$ from its MRPT index set of trees $\mathcal{T}$ and then transmits the set \textit{S} and partition $\mathbf{q}_i$ of $\mathbf{q}$ to worker node $\textit{W}_i$. For every \textit{S}, each worker node $\textit{W}_i$ only fetches the matrix $\mathbf{X}_i(S)^T$ from $\mathbf{X}^T_i$ already stored in its memory. It then computes the product $\mathbf{X}_i(S)^T\mathbf{q}_i$ and returns the resulting vector to the master node. The master node can thus compute the vector $\mathbf{w}$ by adding the results using \eqref{uncoded_equation}, and determine the $k$ nearest neighbors using the exact distances. 

\subsection{Coded Distributed Matrix-Vector Multiplication using MatDot Codes}\label{matdotcode}
In order to successfully compute the vector $\mathbf{w}$ using the strategy described in Section~\ref{uncoded}, the master node must wait for every worker node $\textit{W}_i$ to successfully return the product $\mathbf{X}_i(S)^T\mathbf{q}_i$. In a straggler-prone environment, this might cause unprecedented delays in computation. Thus, to avoid waiting for all nodes and be able to recover the matrix-vector product by only waiting for some out of all workers to finish, we will now apply the MatDot-based distributed matrix multiplication strategy~\cite{fahim2017optimal}. 

We partition the matrix $\mathbf{X}^T$ vertically again, but into $m$ partitions instead of $P$ as follows:
\begin{equation}
\mathbf{X}^T=\begin{bmatrix}\mathbf{X}^T_1&\mathbf{X}^T_2 & \ldots & \mathbf{X}^T_m
\end{bmatrix}.
\end{equation}
We then use the following encoding polynomial:
\begin{equation}\label{mdc_datapoly}
P_{\mathbf{X}^T}(\beta) = \sum_{j=1}^{m}{\mathbf{X}}_j^T{\beta}^{j-1}.
\end{equation}
The rows of $P_{\mathbf{X}^T}(\beta)$ indexed in set $S$ actually represent the following polynomial:
\begin{equation}
P_{\mathbf{X}(S)^T}(\beta) = \sum_{j=1}^{\textit{m}}{\mathbf{X}_j(S)^T{\beta}^{j-1}}.
\label{eq:subset_encoding}
\end{equation}
We will be referring to this observation later.

\begin{figure}
\smallskip
\centering
\includegraphics[height=7.2cm]{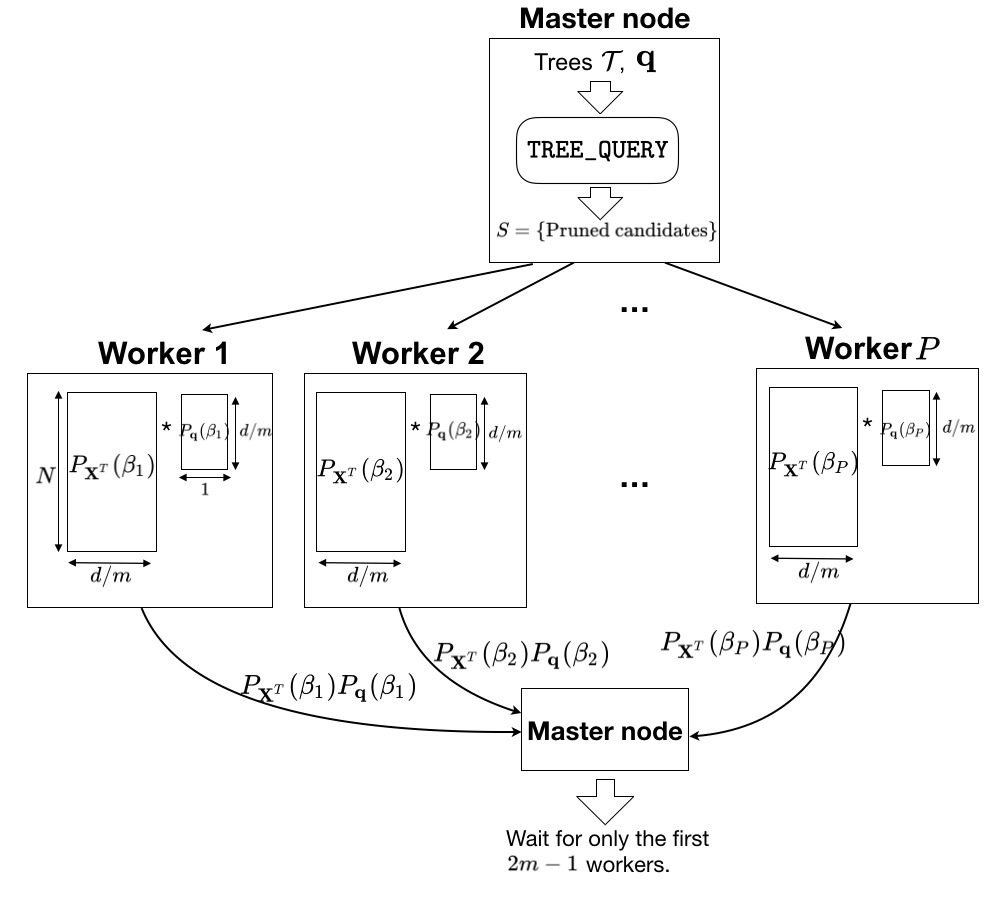}
\caption{Model Parallel Architecture with MatDot Codes.}
\label{mdc_arch}
\end{figure}

Now, given a cluster with a master node and $P$ worker nodes, as shown in Fig. \ref{mdc_arch}, each worker node $\textit{W}_i$ is initialized with a different $\beta_i$, using which it computes the polynomial $\textit{P}_{\mathbf{X}^T}$($\beta_i$) in (\ref{mdc_datapoly}). 
This encoding step can be performed off-line as $\mathbf{X}^T$ is known in advance.

During the online stage, given a query $\mathbf{q}$ for which the \textit{k} nearest neighbors must be determined, the master node first partitions $\mathbf{q}$ into $m$ parts: $\{\mathbf{q}_j : \textit{j} \in \{1,2,\ldots,\textit{m} \} \}$. We then use the following encoding polynomial:
\begin{equation}\label{mdc_querypoly}
P_\mathbf{q}(\beta) = \sum_{j=1}^{\textit{m}}{\mathbf{q}_j{\beta}^{m-j}}.
\end{equation}

As in Section~\ref{uncoded}, the master node first determines the candidate set \textit{S}. It then transmits \textit{S} and the encoded query $P_{\mathbf{q}}(\beta_i)$ obtained from (\ref{mdc_querypoly}) to worker $\textit{W}_i$. The worker $\textit{W}_i$ then fetches only the matrix $P_{\mathbf{X}(S)^T}(\beta_i)$ from its stored $P_{\mathbf{X}^T}(\beta_i)$ (recall \eqref{eq:subset_encoding}) which essentially denotes all the rows of $P_{\mathbf{X}^T}(\beta_i)$ indexed in $S$. Then, it computes the product $P_{\mathbf{X}(S)^T}(\beta_i)P_{\mathbf{q}}(\beta_i)$ and returns the result to the master node.

The coefficient of $\beta^{m-1}$ in the polynomial  $P_{\mathbf{X}(S)^T}(\beta) P_{\mathbf{q}}(\beta)$ turns out to be our desired desired matrix-vector product $\mathbf{X}(S)^T\mathbf{q}= \sum_{j=1}^m \mathbf{X}_j(S)^T\mathbf{q}_j$ from the property of MatDot codes. 
We need to evaluate the polynomial at only $2m-1$ distinct points so as to determine the coefficient for every power of $\beta$. The master node must therefore wait for at least $2m-1$ worker nodes following which it can determine the term $\sum_{i=1}^{m}\mathbf{X}_i(S)^T\mathbf{q}_i$ using polynomial interpolation. 
We then follow the strategy of comparing the exact distances in Section~\ref{new_sec} to obtain the set of the \textit{k} nearest neighbors to $\mathbf{q}$.

\subsection{Coded Distributed Matrix-Vector Multiplication using Systematic MatDot Codes}
In our prior work\cite{fahim2017optimal,dutta2018optimal} where we proposed MatDot Codes, we also introduced their systematic variant. Their advantage is that, while for MatDot Codes the recovery threshold is always $2m-1$, for systematic MatDot Codes one might sometimes only need $m$ nodes to finish, although $2m-1$ is the worst-case value. In this section, we apply the systematic MatDot code to the MRPT problem. 

Similar to the previous case, we first partition $\mathbf{X}^T$ vertically into $m$ partitions, but then use a different encoding function:
\begin{equation}\label{smdc_encoding}
P_{\mathbf{X}^T}(\beta) = \sum_{j=1}^{m}\mathbf{X}^T_{j}L_j(\beta),
\end{equation}
where $L_j(\beta)$ is given by:
\begin{equation}\label{li}
L_j(\beta) = \prod_{r \in \{1,2,\ldots,m\} \backslash j}{\dfrac{\beta-\beta_r}{\beta_j-\beta_r}}.
\end{equation}
Consider a cluster consisting of one master node and $P$ worker nodes. Worker $\textit{W}_i$ is assigned a value $\textit{$\beta$}_i$ using which it computes the polynomial in (\ref{smdc_encoding}). This encoding is performed in advance, in the off-line stage. Interestingly, the workers $\{W_i : i = 1,2,\ldots,m\}$ turn out to be the \textit{systematic} worker nodes, which contain uncoded partitions of $\mathbf{X}^T$. 

In the online phase, given a query $\mathbf{q}$, the master node first partitions $\mathbf{q}$ into $m$ partitions and then uses the following encoding function:
\begin{equation}
P_\mathbf{q}(\beta) = \sum_{j=1}^{m}{\mathbf{q}_{j}L_j(\beta)},
\end{equation}
where $L_j(\beta)$ is given by (\ref{li}).
The master node then transmits candidate set $S$ and $P_\mathbf{q}(\beta_i)$ to each worker node. Worker $\textit{W}_i$ is responsible for computing the product \textit{$P_{\mathbf{X}(S)^T}(\beta_i)P_\mathbf{q}(\beta_i)$}. We first consider the case when the first \textit{m} workers to successfully complete their computation are the \textit{systematic} worker nodes. We can then obtain the vector $\mathbf{w}$ as follows:	
\begin{equation}\label{smdc_final}
\mathbf{w} = \sum_{i=1}^{m}{P_{\mathbf{X}(S)^T}(\beta_i)P_\mathbf{q}(\beta_i)}.
\end{equation}
If the results of the first $2m-1$ successful workers do not contain results from the $m$ systematic nodes, then the master interpolates the polynomial \textit{$P_{\mathbf{X}(S)^T}(\beta)P_\mathbf{q}(\beta)$}. It then computes this polynomial product at each \textit{$\beta_i$} $\in$ \textit{\{$\beta_1$,$\beta_2$,\ldots,$\beta_m$\}}. Finally, it computes the vector $\mathbf{w}$ using (\ref{smdc_final}). Note that in the ideal case, \textit{i.e.}, when all the systematic worker nodes finish first, we only needed $m$ nodes to finish as opposed to the worst-case recovery threshold of $2m-1$. We can now proceed with the steps of comparing exact distances (see Section~\ref{new_sec}) to retrieve the \textit{k} nearest neighbors of $\mathbf{q}$.

\section{EXPERIMENTAL RESULTS}\label{expt}

In this section, we evaluate the effectiveness of data and model parallel MRPT in terms of both accuracy and speed. All of our experiments were conducted on Amazon Elastic Compute Cloud instances~\cite{AmazonAWS}. 

The STL-10 dataset~\cite{coates2011analysis} is a dataset of $N=100000$ images each of dimension $d=9216$ used in unsupervised image classification algorithms, while GIST~\cite{jegou2011product} is a popular dataset with $N=1000000$ and $d=960$ used in ANN algorithms. These datasets provide us with a good mix of dimensionality and number of datapoints to evaluate our proposed strategies. The MRPT parameters used for the experiments are provided in Table \ref{mrpt_params}.
\begin{table}[h]
\caption{MRPT Parameters}
\label{mrpt_params}
\begin{center}
\begin{tabular}{|c||c||c||c||c|}
\hline
\textit{Dataset} & \textit{$\ell$} & \textit{$\nu$} & \textit{Number of Trees} & \textit{Projection Sparsity}\\
\hline
STL-10 & 7 & 25 & 900 & 0.01\\
GIST & 9 & 10 & 900 & 0.032\\
\hline
\end{tabular}
\end{center}
\end{table}

We evaluate the accuracy of our implementation using recall defined as: $\dfrac{\vert \kappa_{MRPT} \cap \kappa \vert}{k}$, where $\mathbf{q}$ is the query whose true \textit{k} nearest neighbors is the set $\kappa$ and ${\kappa_{MRPT}}$ is the set of \textit{k} nearest neighbors returned by the algorithm.

For the single node MRPT baseline, we used a compute optimized \textit{c5.large} instance with two 3GHz Intel Xeon Platinum processors and 4 GB of memory. For the data parallel and model parallel architectures, we used \textit{t2.medium} instances with two 2.3 GHz Intel Broadwell processors and 4 GB of memory. All instances are provisioned with 40GB HDD secondary storage. Note that the hardware used for our baseline experiments is superior to that used to evaluate our parallel strategies. In all experiments we ran 500 queries sequentially with $k=10$. We consider the average result of 50 runs for each experiment. The experiments were conducted in a cluster consisting of 1 master node and 16 worker nodes.

In the experiments with MatDot codes and systematic MatDot codes, we used encoding polynomials of degree 2.

\subsection{System Constraints}
All our experiments are conducted on systems with limited memory. Thus the MRPT algorithm has to use the disk to hold the index $\mathcal{T}$ and the actual data points. In comparison, the data parallel and model parallel architectures use less memory and avoid disk penalties. In the data parallel architecture, we reduce the amount of points each worker must hold by a factor of \textit{P}. Additionally, the MRPT indexes become smaller as each worker holds a smaller fraction of the dataset. In the model parallel architecture, the master node can discard the components for each point after tree construction; it only needs the index. At each worker node, the memory requirement is at least halved.

\subsection{Simulating Stragglers on Amazon Web Services}
To demonstrate the effects of stragglers in the model parallel architecture, we sample the minimum time $T_i$ a worker $W_i$ must take to complete a matrix multiplication for a query from the shifted Exponential~\cite{lee2018speeding,dutta2016short} and Weibull distributions~\cite{dutta2017coded,heterogeneousclusters} shown in (\ref{exp}) and (\ref{weib}) respectively. 
\begin{align}
Pr[T_i\le t] &= 1 - e^{-\frac{\mu_i}{l_i}(t - a_il_i)} \label{exp}\\
Pr[T_i\le t] &= 1 - e^{\left({{-\frac{\mu_i}{l_i}(t - a_il_i)}}\right)^{\alpha_i}}
\label{weib}
\end{align}

Here, $l_i$ is the number of row vectors loaded at worker $W_i$ for matrix multiplication, $a_i > 0$ is the shift parameter, $\alpha_i > 0$ is the shape parameter for the Weibull distribution, $\mu_i > 0$ is the straggling parameter for $W_i$, and $t \ge {a_i}{l_i}$. For our experiments, we set each $\mu_i$, $a_i$, and $\alpha_i$ to some constants $\mu$, $a$, and $\alpha$ to maintain homogeneity in minimum computation times across workers. The parameters chosen are as follows:
Shifted Exponential ($a=0.0000001$, $\mu=15$) and
Weibull ($a=0.2$, $\mu=2$, $\alpha=0.5$) for both the datasets.
Note that, for model parallel MRPT, $l_i = \vert S\vert$. We use a similar strategy to simulate straggling in data parallel MRPT with $l_i$ set to the average of $\vert S\vert$ for the set of test queries. 



\subsection{Results}
\begin{figure*}[htbp!]
    \centering 
\begin{subfigure}{0.48\linewidth}
\includegraphics[height=4.5cm]{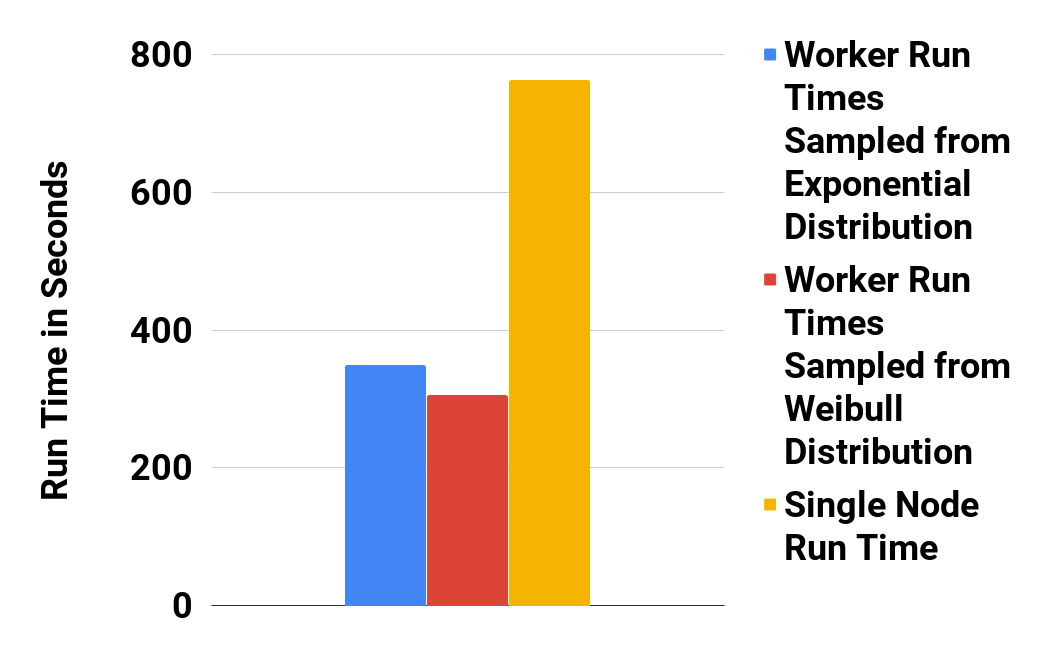}
\caption{Execution Times: Single Node \& Data Parallel (STL-10).}
\label{result_data_parallel_stl}
\end{subfigure}\hfil 
\begin{subfigure}{0.47\linewidth}
\includegraphics[height=4.5cm]{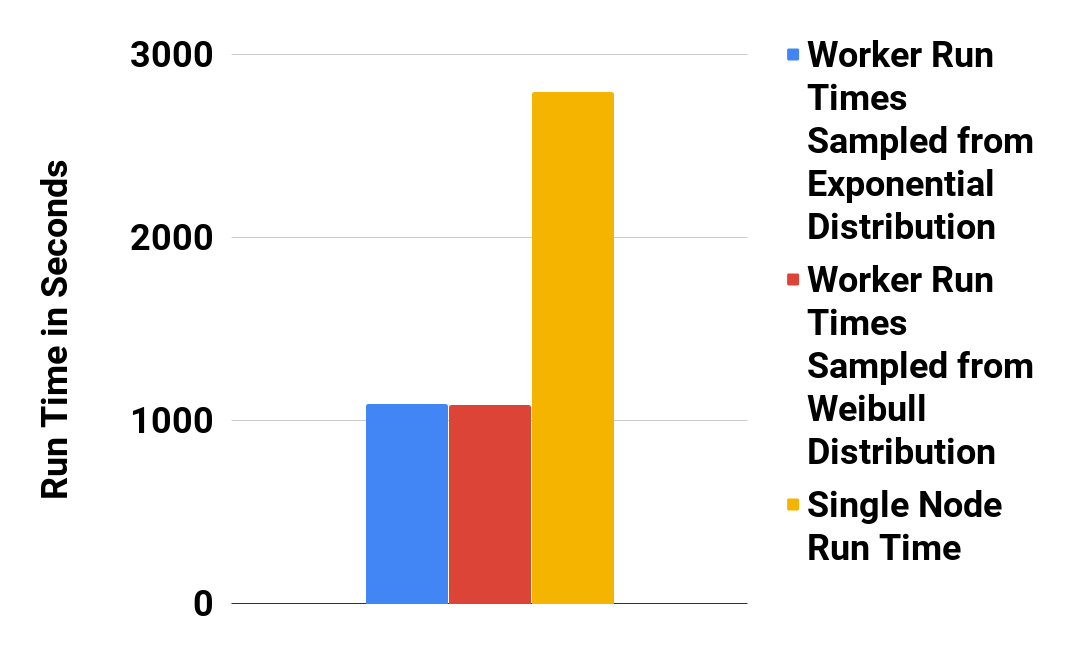}
\caption{Execution Times: Single Node \& Data Parallel (GIST).}
\label{result_data_parallel_random}
\end{subfigure}\hfil 

\medskip
\begin{subfigure}{0.49\linewidth}
\includegraphics[height=3.5cm]{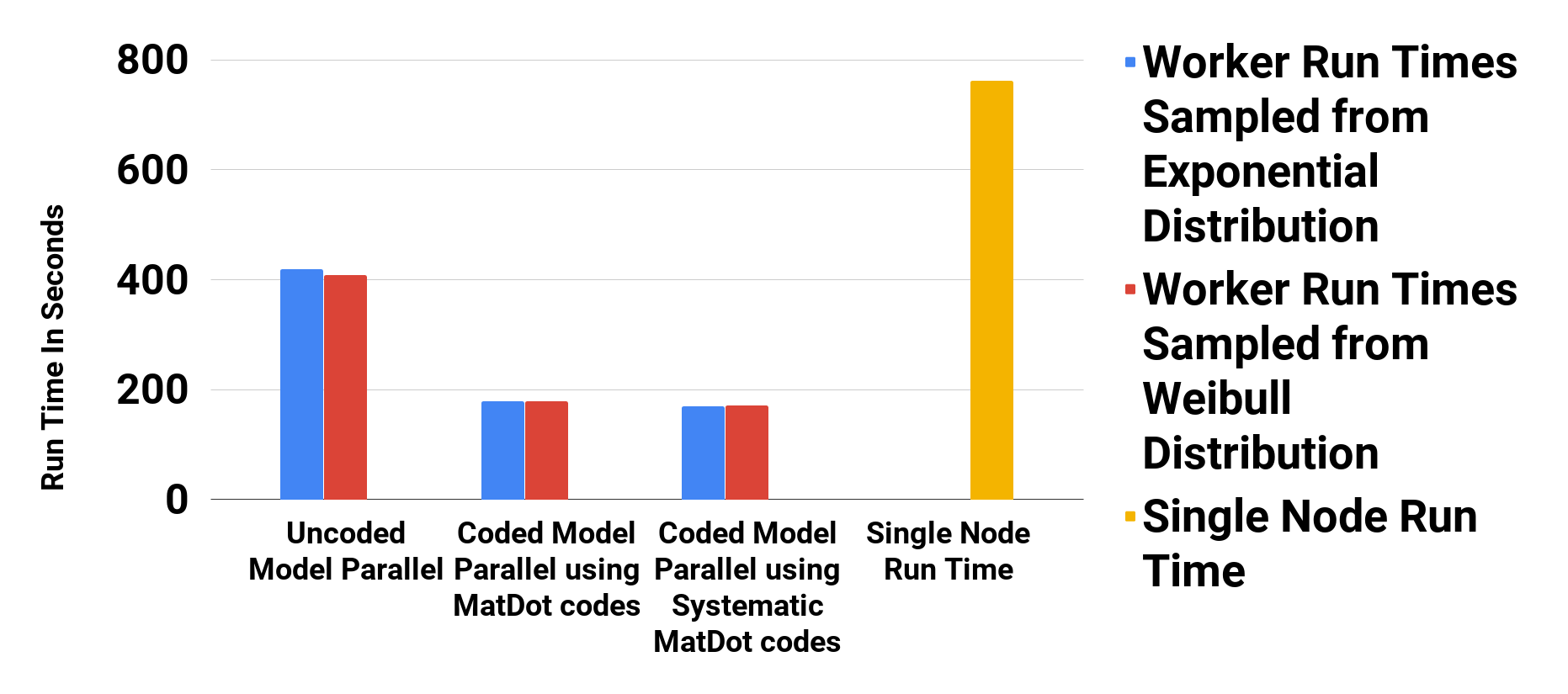}
\caption{Execution Times: Single Node \& Model Parallel (STL-10).}
\label{result_model_parallel_stl}
\end{subfigure}
\begin{subfigure}{0.47\linewidth}
\includegraphics[height=3.5cm]{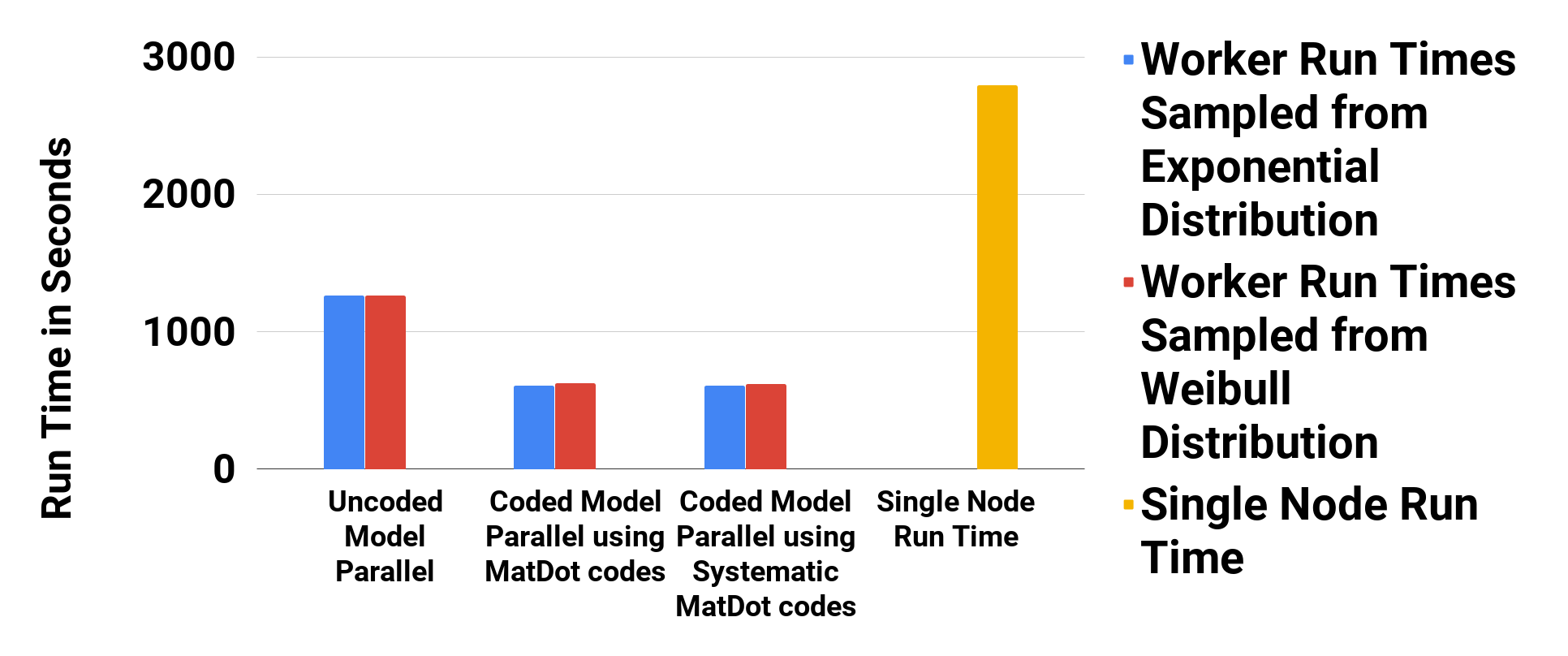}
\caption{Execution Times: Single Node \& Model Parallel (GIST).}
\label{result_model_parallel_random}
\end{subfigure}\hfil 

\medskip
\begin{subfigure}{0.47\linewidth}
\includegraphics[height=4cm]{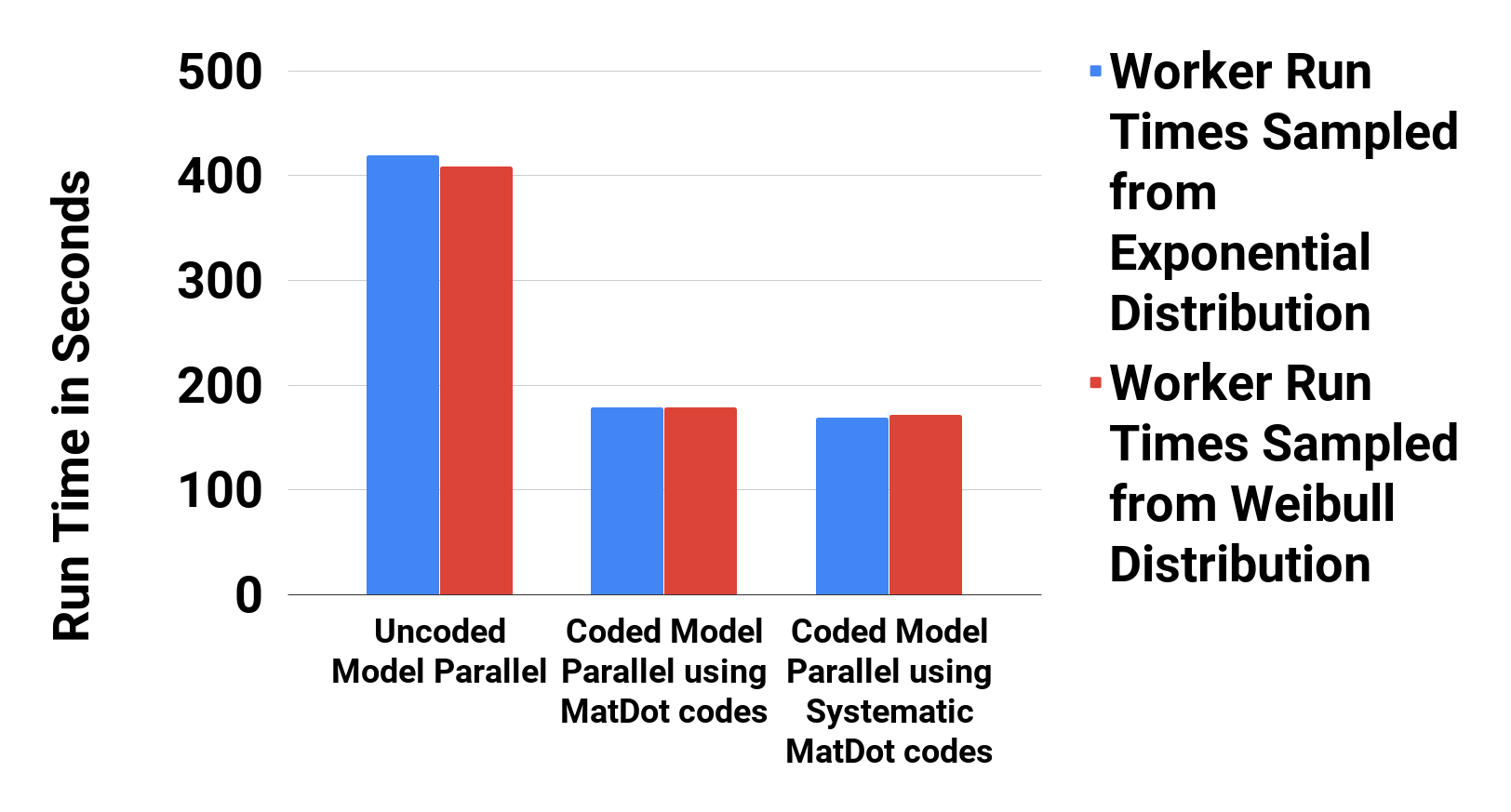}
\caption{Effect of Coded Matrix Multiplication for STL-10.}
\label{result_uncoded_vs_coded_stl}
\end{subfigure}\hfil 
\begin{subfigure}{0.47\linewidth}
\includegraphics[height=4cm]{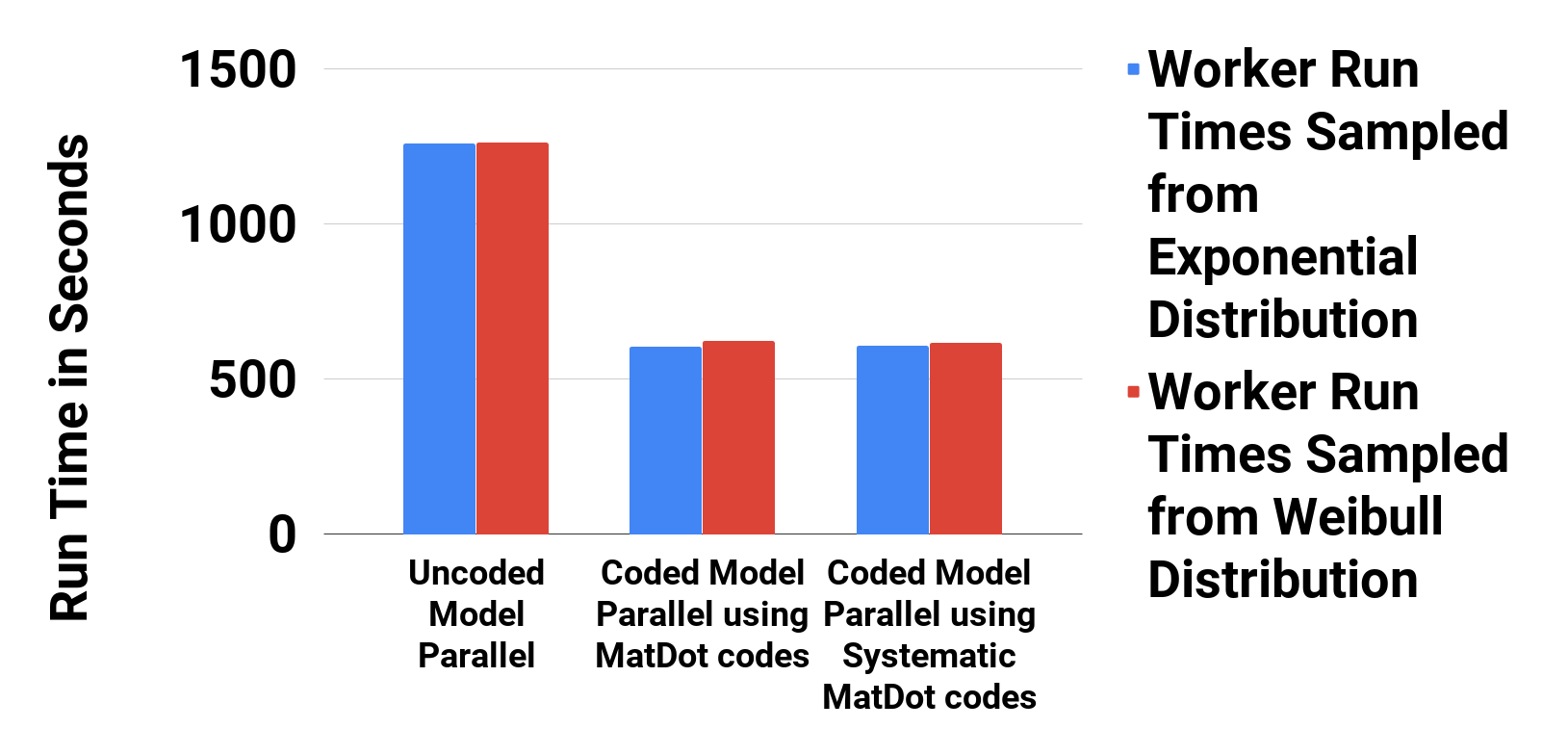}
\caption{Effect of Coded Matrix Multiplication for GIST.}
\label{result_uncoded_vs_coded_random}
\end{subfigure}
\caption{Experimental Results of MRPT in different configurations.}
\label{experiments}
\end{figure*}

Our experimental results are provided in Tables~\ref{stats} and also illustrated in Fig.~\ref{experiments}. For completion, we also include our obtained recall values here:\\
STL-10: Data Parallel (0.9632), Model Parallel (0.9648).\\
GIST: Data Parallel (0.9430), Model Parallel (0.9350).

\begin{table*}[h]
\caption{Runtime Statistics in seconds for MRPT on STL-10 and GIST datasets over 50 Runs}
\label{stats}
\begin{center}
\begin{tabular}{|l|c|c|c|c|}
\hline
\textit{Experiment Type} & \textit{Mean (STL-10)} & \textit{Std. Dev.(STL-10)}& \textit{Mean (GIST)} & \textit{Std. Dev.(GIST)} \\
\hline
Single Node & 762.272 & 69.147& 2795.791 & 173.395 \\
Data Parallel, Shifted-Exponential Runtime & 349.309 & 14.672& 1091.010 & 44.261 \\
Data Parallel, Weibull Runtime & 305.046 & 14.44 & 1082.638 & 39.081 \\
Uncoded Model Parallel, Shifted-Exponential Runtime & 419.114 & 17.217 & 1260.504 & 49.566 \\
Uncoded Model Parallel, Weibull Runtime & 408.814 & 13.625 & 1261.980 & 62.259 \\
Coded Model Parallel (MatDot), Shifted-Exponential Runtime & 179.082 & 7.956 & 608.496 & 38.406 \\
Coded Model Parallel (MatDot), Weibull Runtime & 179.075 & 8.312 & 618.515 & 39.270 \\
Coded Model Parallel  (Systematic MatDot), Shifted-Exponential& 168.869 & 6.865  & 604.546 & 22.951 \\
Coded Model Parallel  (Systematic MatDot), Weibull Runtime &171.428 & 8.526 & 622.716 & 22.495 \\
\hline
\end{tabular}
\end{center}
\end{table*}
In all our experiments, both data and model parallel MRPT outperform the single node implementation. As shown in Fig. \ref{result_data_parallel_stl} and Fig. \ref{result_data_parallel_random}, data parallel MRPT is significantly faster than single node MRPT. This is due to the smaller size of the MRPT index and absence of disk penalties at workers. It can be seen that the data parallel strategy has better performance when compared to uncoded model parallel strategy because of its embarrassingly parallel design. Owing to this design, the data parallel strategy could scale linearly with the number of nodes. However, these scaling benefits in query execution come at a cost, as the random projection trees computed at each node do not contain all the data points and hence the candidate set generated by each node could contain lesser true positives. To offset this condition, we might have to lower the voting threshold in-order to generate a better candidate, while causing more communication overheads and hence lower query execution time. The note mentioned in Section III explains this case in more detail. For our experiments, we do not lower the voting threshold as the loss in recall for STL-10 and GIST is not significant.
	
The model parallel architecture results in a high communication cost as the candidate set has to be transmitted to each worker node. However, we find that if the MRPT parameters for a dataset are sufficiently tuned, Algorithm~\ref{mrpt_algorithm} will generate a smaller candidate set over which exact distance calculations must be performed. For our experiments, the algorithm was able to reduce the search space for STL-10 from 100000 datapoints to 3025 and for GIST from 1000000 to 14934 datapoints on average per query. Additionally, the model parallel strategy does not suffer from the accuracy related issues as compared to the data parallel architecture as the candidate set is generated using the entire dataset and not parts of it separately.

Both the strategies outperform baseline single node MRPT \emph{despite running on inferior hardware}. The parallel strategies are not limited by memory constraints as in the case of single node MRPT. These strategies are therefore very useful when large datasets do not fit in memory. The coded model parallel strategy also makes the algorithm tolerant to slow nodes and failures in a distributed setting.



Fig. \ref{result_model_parallel_stl} and Fig. \ref{result_model_parallel_random} show that model parallel MRPT outperforms single node MRPT. They also outperform data parallel MRPT under simulated straggling. This is of significance to real world systems where straggling may manifest as unreliable nodes or network delays.
Fig. \ref{result_uncoded_vs_coded_stl} and Fig. \ref{result_uncoded_vs_coded_random} show the benefits of coded matrix multiplication as opposed to the uncoded model parallel architecture  under simulated straggling. Both MatDot codes and systematic MatDot codes are consistently faster than the uncoded approach. Fig. \ref{result_uncoded_vs_coded_stl} also shows that systematic MatDot codes is able to outperform MatDot codes owing to its lower recovery threshold. 




\section{CONCLUSIONS}
\label{sec:conclusion}
We proposed two approaches to parallelize the MRPT algorithm in a distributed setting. We also applied the MatDot code based distributed matrix multiplication strategy to reduce the recovery threshold in a system that is prone to stragglers. We showed that our parallelization strategies can achieve faster queries than the single node MRPT algorithm under limited memory. Our results demonstrate the benefits of applying MatDot code and systematic MatDot code to the model parallel architecture in a system with simulated stragglers. In our experiments we observed large floating point errors when inverting high degree Vandermonde matrices for polynomial interpolation. As future work, we will experiment with strategies to reduce the condition number of a Vandermonde matrix~\cite{tyrtyshnikov1994bad} so that we can employ polynomials of higher degrees, and thus apply the MatDot code and systematic MatDot code with a larger number of worker nodes. Another possibility is to perform the computations in exact rational arithmetic. This would eliminate rounding errors, but the effect on runtime needs to be analyzed.

\section*{Acknowledgements} This work was supported by
NSF CNS-1702694, the Academy of Finland under the WiFIUS program, the Academy of Finland COIN CoE and NSF CCF 1350314.

\bibliographystyle{IEEEtran}
\bibliography{IEEEabrv,refs}

\end{document}